          [2001/04/25 1.1 (PWD)]
\documentclass[a4paper,twocolumn]{esapub} 

\usepackage{natbib}
\usepackage{graphicx}

\title{The Sample of Gamma-ray Bursts Observed With SPI-ACS}
\author{Arne Rau}
\author{A. von Kienlin}
\affil{Max-Planck-Institut f\"ur Extraterrestrische Physik, 85741 Garching, Germany}
\author{K. Hurley}
\affil{UC Berkeley Space Sciences Laboratory, Berkeley, CA 94720-7450, USA}
\author[1]{G. G. Lichti}

\def\etal{{et\,al. }}

\begin{document}


\maketitle

\begin{abstract}
The SPI anticoincidence shield consists of 91 BGO crystals and is
operated as a nearly omnidirectional gamma-ray burst detector above
~75 keV. Since the start of the mission 269 gamma-ray burst candidates
have been detected. 110 bursts have been confirmed with the
instruments included in the 3rd Interplanetary Network. Here we
present a preliminary statistical analysis of the SPI-ACS sample of
gamma-ray bursts and gamma-ray burst candidates; in particular we
discuss the duration distribution of the bursts. A prominent
population of short burst candidates (duration $<$200\,ms) is found
which is discovered to be strongly contaminated by cosmic-ray nuclei
interacting in the detectors.
\end{abstract}

\section{Introduction}

The detection and investigation of cosmic gamma-ray bursts (GRBs) is
one of the important scientific objectives of the INTEGRAL
mission. Although discovered more than three decades ago (Klebesadel
\etal 1973) the GRB phenomenon is still challenging,
with many open questions to solve. It has long been known that there
are two distinct classes of GRBs which differ observationally in
duration and spectral properties. This was quantified using data from
the Burst and Transient Source Experiment (BATSE) detectors (Fishman
\etal 1989) aboard NASA's Compton Gamma-ray Observatory (CGRO). The
sample of more than one thousand bursts included in the 4$^{th}$ BATSE
catalogue displayed a strong bimodal distribution, with one group
having short durations ($<$1\,s) and hard energy spectra, while the
other was of longer duration (seconds to minutes) with spectra peaking
at softer energies.

We know now that at least some of the long-duration GRBs mark the
deaths of massive stars at cosmological distances. The cosmological
nature of GRBs was demonstrated with the detection of the first
afterglows by BeppoSAX and the measurement of redshifts (Costa \etal
1997). The small sample of bursts with measured redshifts ($\sim$30)
presently extends from z=0.0085 to z=4.511. Observational proof that
GRBs are connected with hypernovae and thus with the final stages of
the evolution of massive stars comes from the detection of supernova
signatures in the afterglow spectra of GRB 030329 (Hjorth
\etal 2003, Stanek \etal 2003).

The INTEGRAL mission contributes to GRB science in two ways. (i) For
bursts which occur in the field of view (FoV) of the Spectrometer SPI
and of the imager (IBIS), INTEGRAL provides accurate positions for
rapid ground and space-based observations. Also high energy spectra in
the range of 20\,keV to 8\,MeV are provided (see von Kienlin \etal
2003 for a summary). (ii) The anticoincidence shield of SPI (SPI-ACS)
acts as an omnidirectional GRB detector and produces time profiles
with a resolution of 50\,ms.

\section{Gamma-ray burst Detection Capabilities of SPI-ACS}

The anticoincidence shield consists of 91 Bi$_4$Ge$_3$O$_{12}$ (BGO)
crystals with a total mass of 512\,kg. It provides a large
($\sim$0.3\,m$^2$) effective area for the detection of gamma-ray
bursts. Each of the crystals is viewed redundantly by two
photomultipliers and only the total event rate of all the crystals is
telemetered to Earth with 50\,ms time resolution. Therefore, the
SPI-ACS setup does not provide any spatial or spectral
resolution. Unfortunately, the energy range is not exactly known
either, as the thresholds of the individual photomultipliers are
different due to different light yields of the crystals. The lower
energy threshold can only be estimated to be $\sim$75\,keV. Nothing
accurate can be said about the upper end of the energy range, except
that it is above 10\,MeV (for more details see von Kienlin \etal
2003). As the SPI-ACS surrounds the spectrometer nearly completely, it
provides a quasi-omnidirectional field of view.

SPI-ACS is part of the INTEGRAL Burst Alert System (IBAS; see
Mereghetti 2004, these proceedings). A software trigger algorithm
searches for an excess in the overall count rate with respect to a
running average background. For each trigger an ASCII light curve (5\,s
pre-trigger to 100\,s post-trigger) and the spacecraft ephemeris
are stored and made publicly
available\footnote{http://isdcarc.unige.ch/arc/FTP/ibas/spiacs/}. 

Since December 2002 SPI-ACS has been an important member of the
3$^{rd}$ Interplanetary Network (IPN) of $\gamma$-ray burst detectors,
which provides burst localizations using the triangulation method
(Hurley 1997; Hurley \etal 2004, these proceedings). By studying
triangulations of bursts with accurately known positions such as
Soft-Gamma-ray Repeaters and GRBs with counterpart detections, the
absolute timing uncertainty of SPI-ACS was estimated to be of the
order of 100\,ms. Recently, additional triangulations of some
bursts with precisely known localizations using Konus-Wind and/or
Helicon-Coronas-F instruments in conjunction with SPI-ACS data
revealed a systematic timing inaccuracy such that the SPI-ACS was
lagging behind UTC by 134$\pm$12\,ms (Golenetskii, Mazets, Pal'shin \&
Frederiks, private communication 2004). We further investigated the
timing error and confirmed a 125$\pm$25\,ms lag by comparing the rise
of (most likely) particle-induced short events in SPI-ACS and
corresponding saturations in the Ge-detectors of SPI (see
below). Starting with events which occurred after April 2 2004 11:00
UTC this time lag of 125\,ms has been corrected for in the publicly
available SPI-ACS data files by a fix in the software. The origin of
this offset is still unknown.

\section{Sample Selection}

A thorough and robust selection is required for a decent statistical
study of the GRBs detected in the SPI-ACS overall count
rate. As SPI-ACS lacks an imaging capability, a direct selection by
localization, which would distinguish particle or solar
flare events, is not feasible. Although localization, or at least the
confirmation of non-solar system origin, can be provided by the
instruments participating in the IPN, the remaining events would form a
conservative sample. A fraction of the real GRBs would escape
selection due to the different sky coverage and energy ranges of
the instruments. In order to analyze the entire sample of GRB
candidates detected with SPI-ACS, we therefore define our selection
criteria independent of the confirmation by other instruments and
base it on the only measurable quantity from the SPI-ACS data, the observed
light curve.

\begin{figure*}
   \centering
   \includegraphics[width=0.35\linewidth,angle=270]{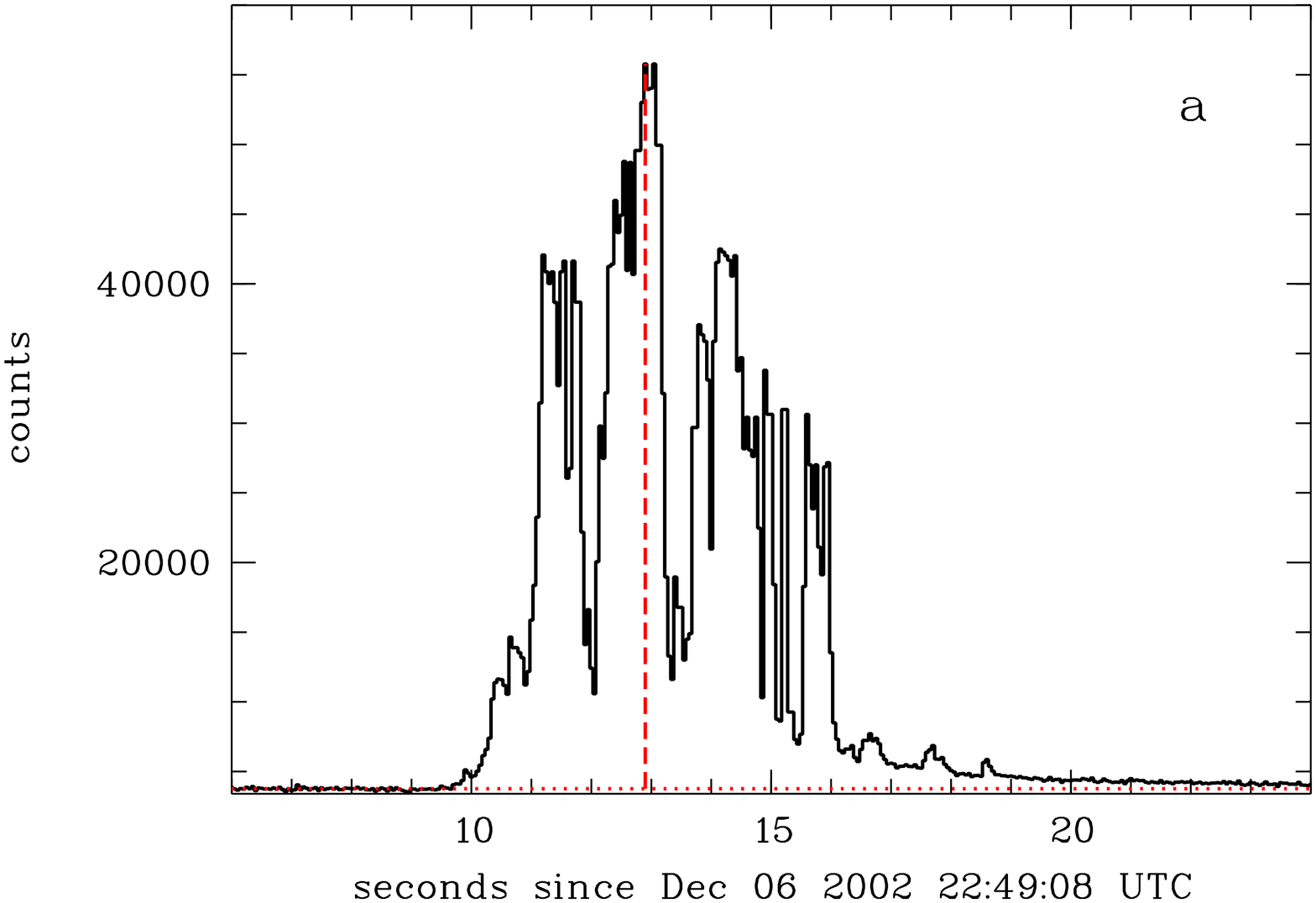}\hspace{-.4cm} 
   \includegraphics[width=0.35\linewidth,angle=270]{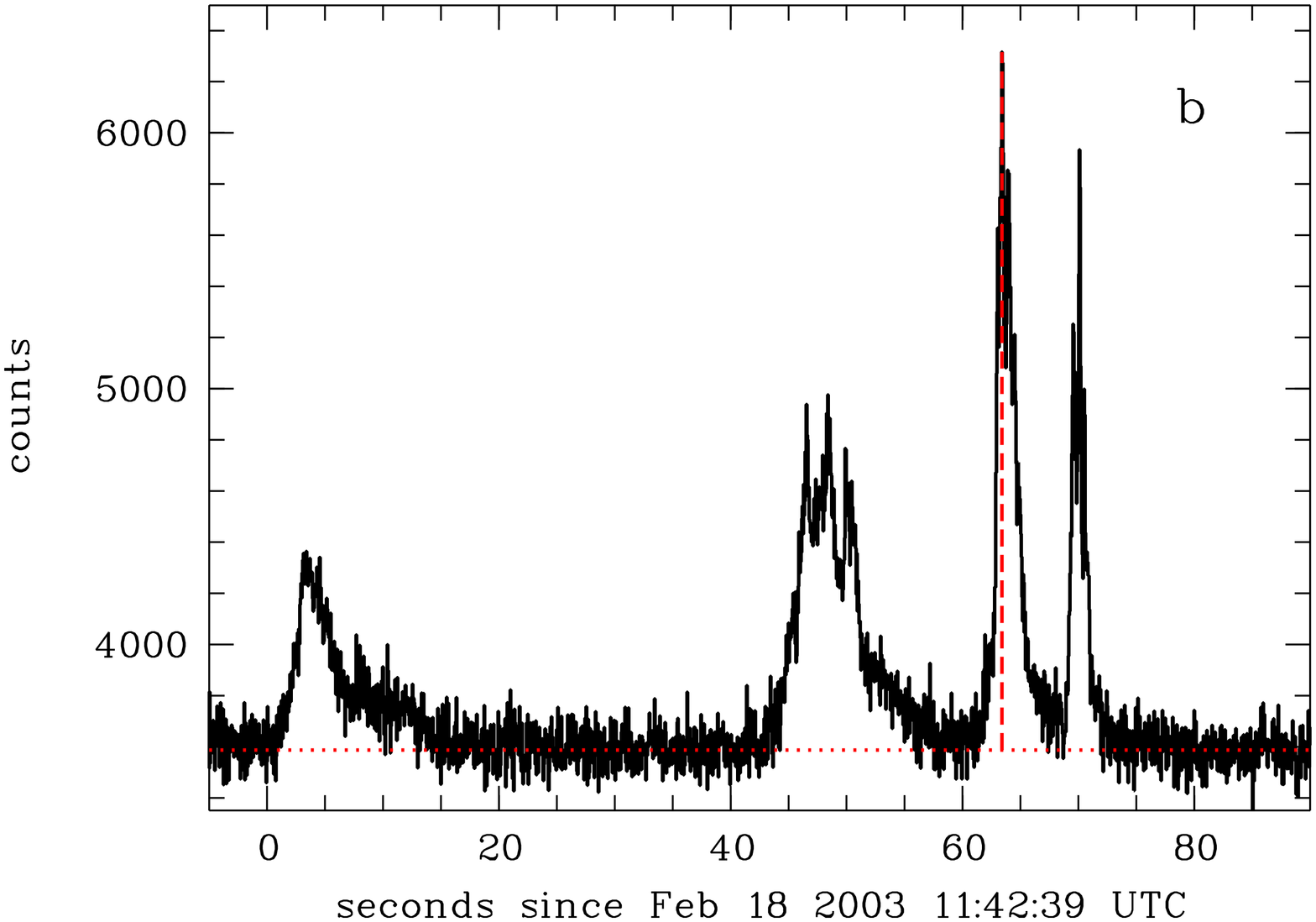} 
   \includegraphics[width=0.35\linewidth,angle=270]{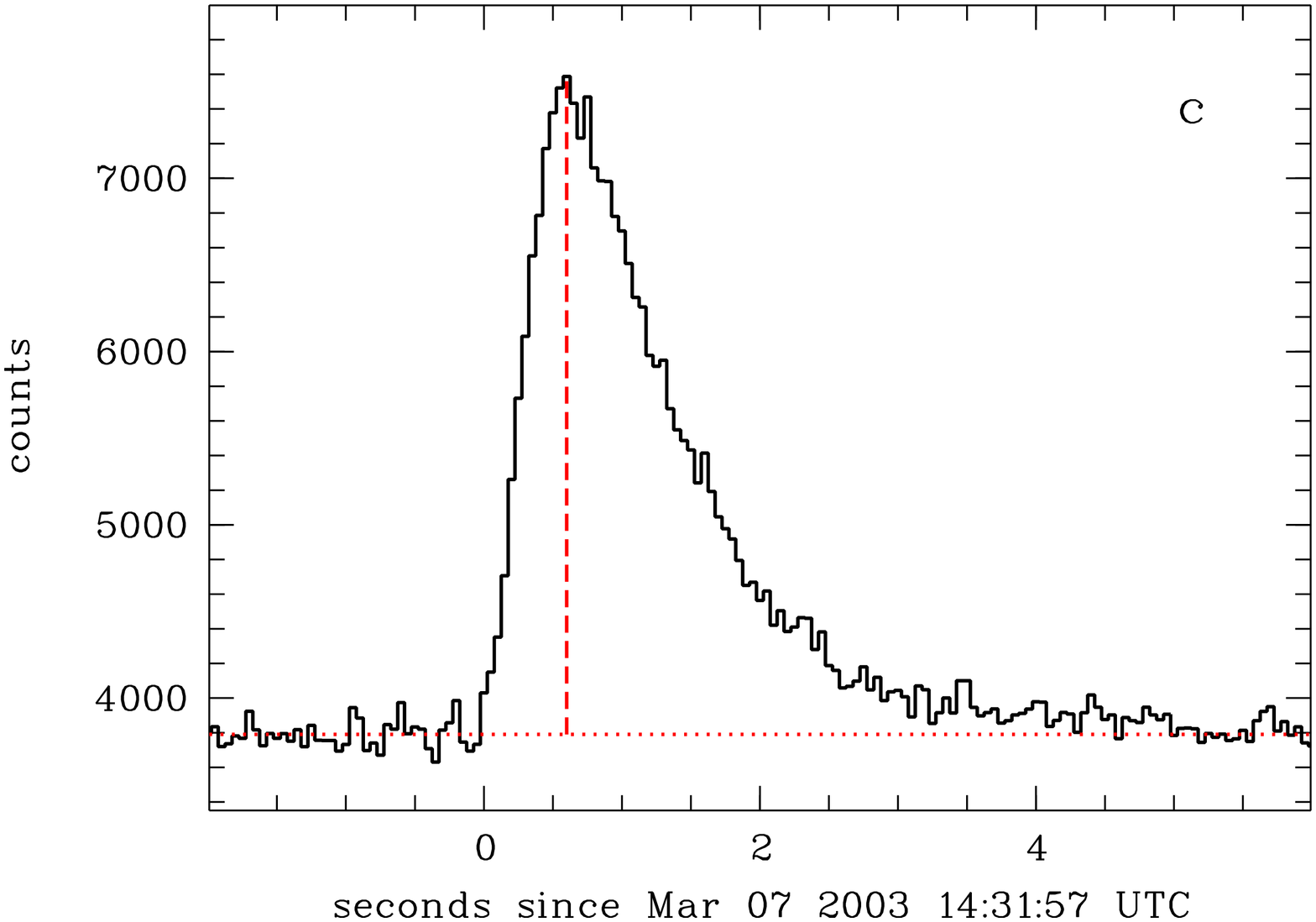}\hspace{-.4cm} 
   \includegraphics[width=0.35\linewidth,angle=270]{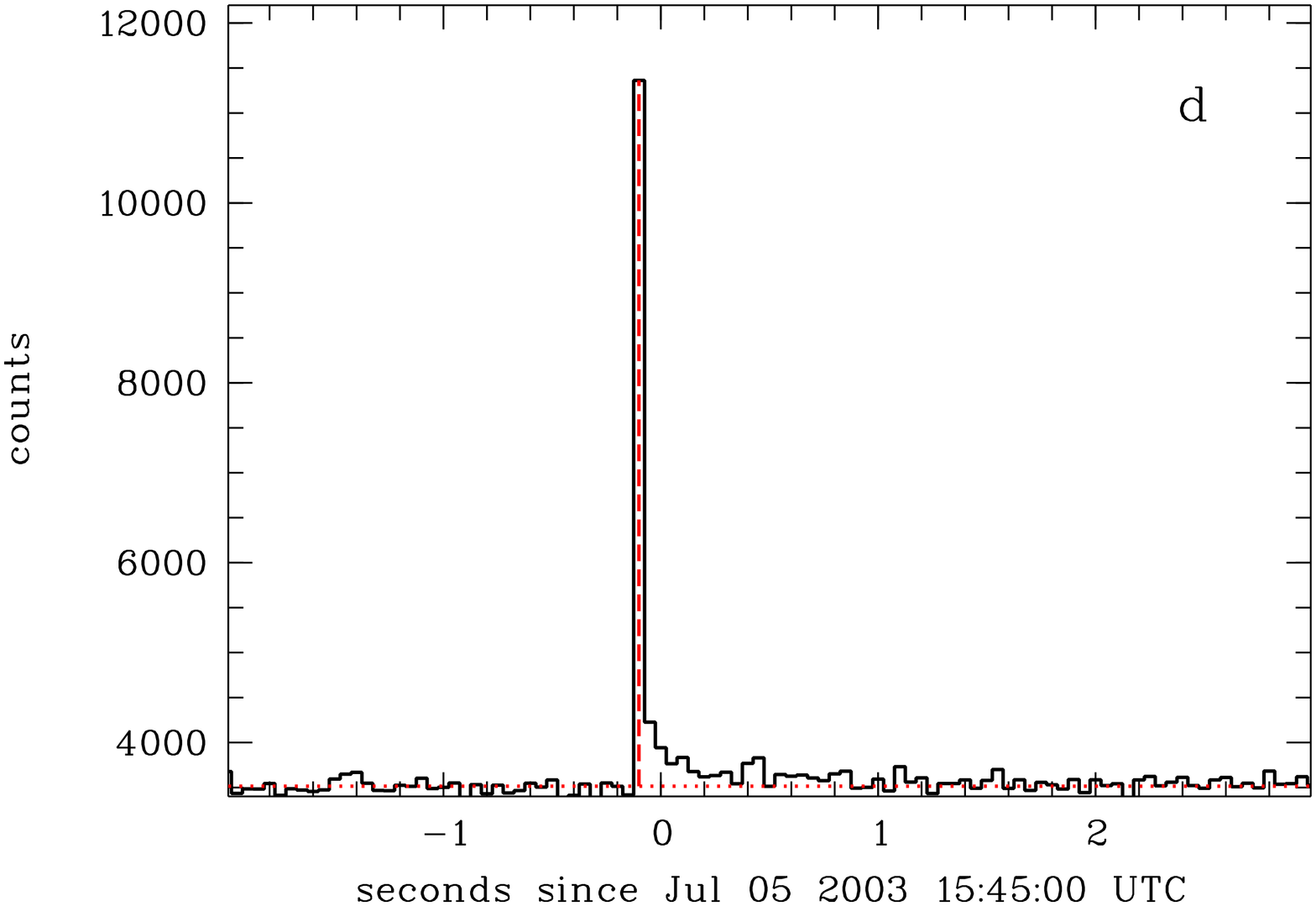} 
   \includegraphics[width=0.35\linewidth,angle=270]{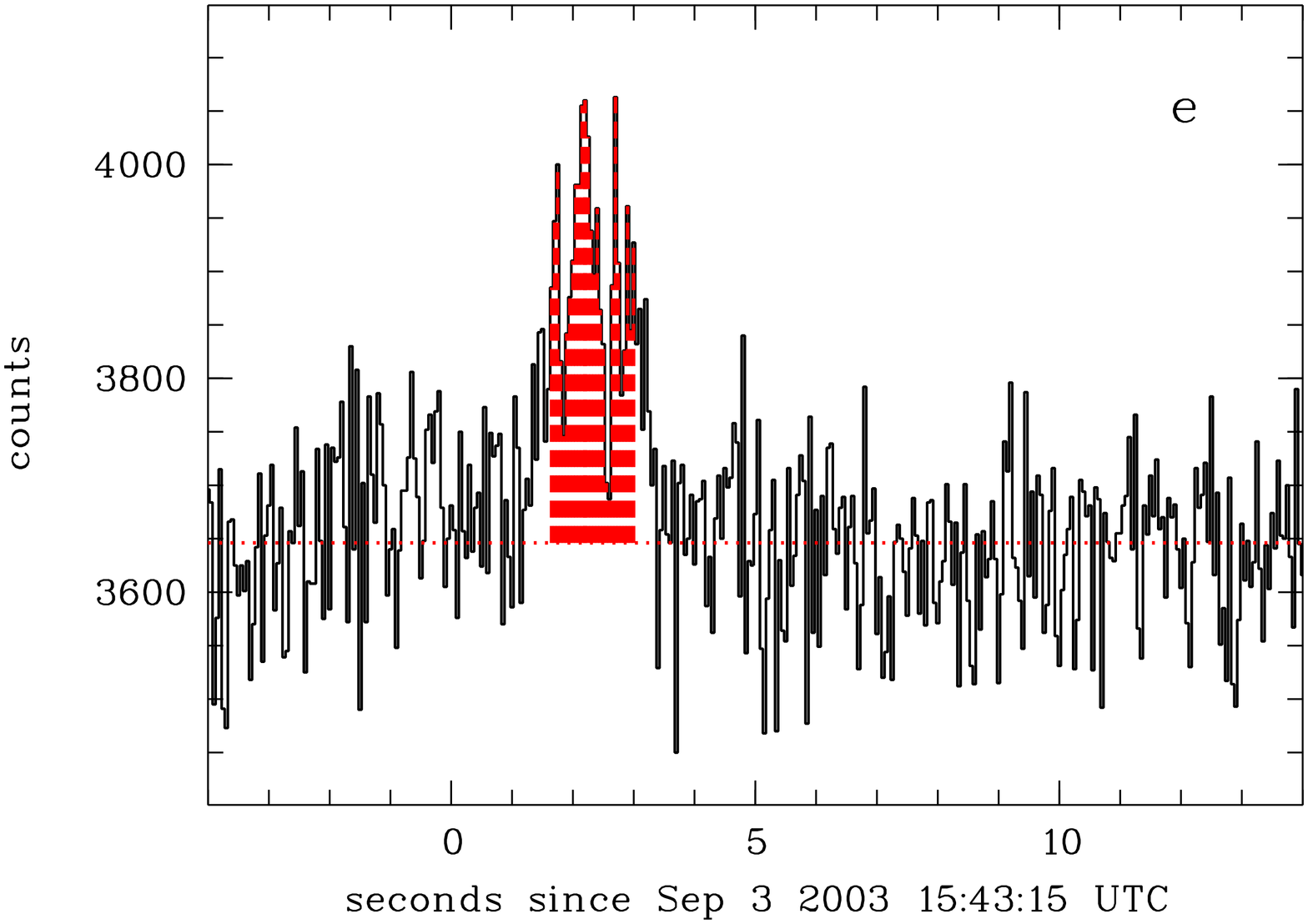}\hspace{-.4cm} 
   \includegraphics[width=0.35\linewidth,angle=270]{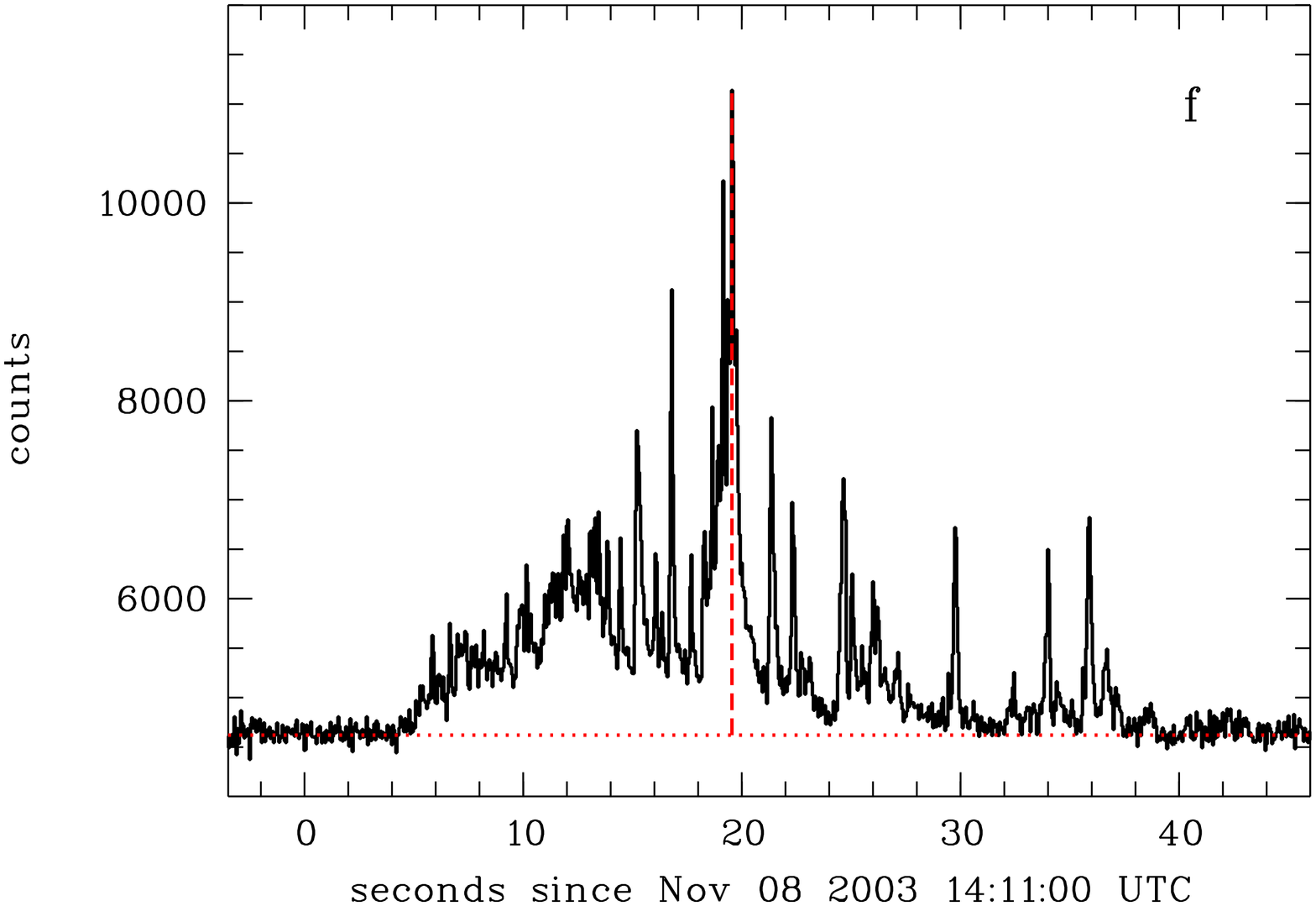}
   \caption{SPI-ACS overall count rate light curves with 50\,ms time
   binning for a selection of GRB candidates from our sample. The
   horizontal dotted lines mark the background and the vertical dashed
   lines show the number of bins sufficient for including the event
   in the GRB sample following the selection criteria described in
   the text. a) the exceptionally bright GRB 021206, triangulated using
   Ulysses, RHESSI, Mars Odyssey and Konus-Wind (Hurley
   \etal 2003c); b) GRB 030218, also seen by Ulysses (Hurley
   \etal 2003a); c) bright, intermediate duration GRB 030307,
   triangulated using Ulysses, Konus-Wind, HETE (FREGATE) and
   RHESSI (Hurley \etal 2003b); d) short event from July 5 2003, 
   observed only by SPI-ACS; e) faint GRB 030903, also observed by
   Ulysses, Mars Odyssey (HEND) and Konus-Wind (Hurley \etal 2003d);
   f) GRB 031108, also observed by Ulysses, Konus-Wind, Mars Odyssey
   (HEND+GRS) and RHESSI (Hurley \etal 2003e).
   \label{fig:lcExamples}} 
   \end{figure*}

An event is included in the sample as a GRB candidate when the total
significance, $S$, above the background, $B$, exceeds $S$=12$\sigma$
in any time interval during the event. Here, 1$\sigma$ corresponds to
1.6$\times\sqrt{B}$, where the factor 1.6 takes into account the
measured deviation of the background noise from a Poissonian
distribution (von Kienlin \etal\ 2003, Ryde \etal 2003). In order to
test an individual event for the selection criterium, the
significance of the 50\,ms bin with the highest count rate is first
calculated. If $S\geq$12$\sigma$ the event is treated as a GRB
candidate and included in the preliminary sample (see
Fig.~\ref{fig:lcExamples}a--d \& f). Otherwise, the integrated
$S$ of the highest count rate bin together with the neighbouring bin
with the next highest count rate is computed and checked against the
threshold. The procedure is continued until $S\geq$12$\sigma$ or the
105\,s SPI-ACS count rate window around the event is entirely tested
without reaching the threshold. In that case the event is
ignored. GRB 030903 (Fig.~\ref{fig:lcExamples}e) is an example of a
faint event in the sample which required the integration of the
significance of adjacent bins. Each event in the primary sample is
subsequently checked for solar or particle origin using JEM-X and
the GOES web page\footnote{http://www.sec.noaa.gov} and IREM,
respectively.

\section{Results}

Using the selection criteria described in the previous section, a
total of 269 GRB candidates were detected during the first 15 months
(Nov 11 2002 -- Feb 11 2004) of the mission. 110 candidates have been
confirmed by detections with other gamma-ray burst instruments.  Taking the
elapsed mission time into account, this gives a rate of 215 candidates and 88
confirmed bursts per year selected according to our criteria. For
comparison, Lichti \etal (2000) predicted the rate of GRBs prior the
start of INTEGRAL to be $\sim$160 per year at a 10$\sigma$ level.

In addition to the number of events, the SPI-ACS overall detector
count rate provides the possibility to derive the burst duration in
the instrumental observer frame and the variability of the light
curve. As no spectral information exist (only events above
$\sim$75\,keV are recorded), typical burst parameters like fluence and
peak flux cannot be derived from the data. Only quantities like the
total integrated counts and the counts at the burst maximum can be
extracted from the light curve. Here we will focus on the
observer frame duration and its distribution and will discuss the
findings. Additional statistical properties are presented elsewhere (Ryde
\etal 2003) and will be extensively discussed in Rau \etal (2004, 
in preparation).

\begin{figure*}
   \centering
   \includegraphics[width=0.6\textwidth,angle=270]{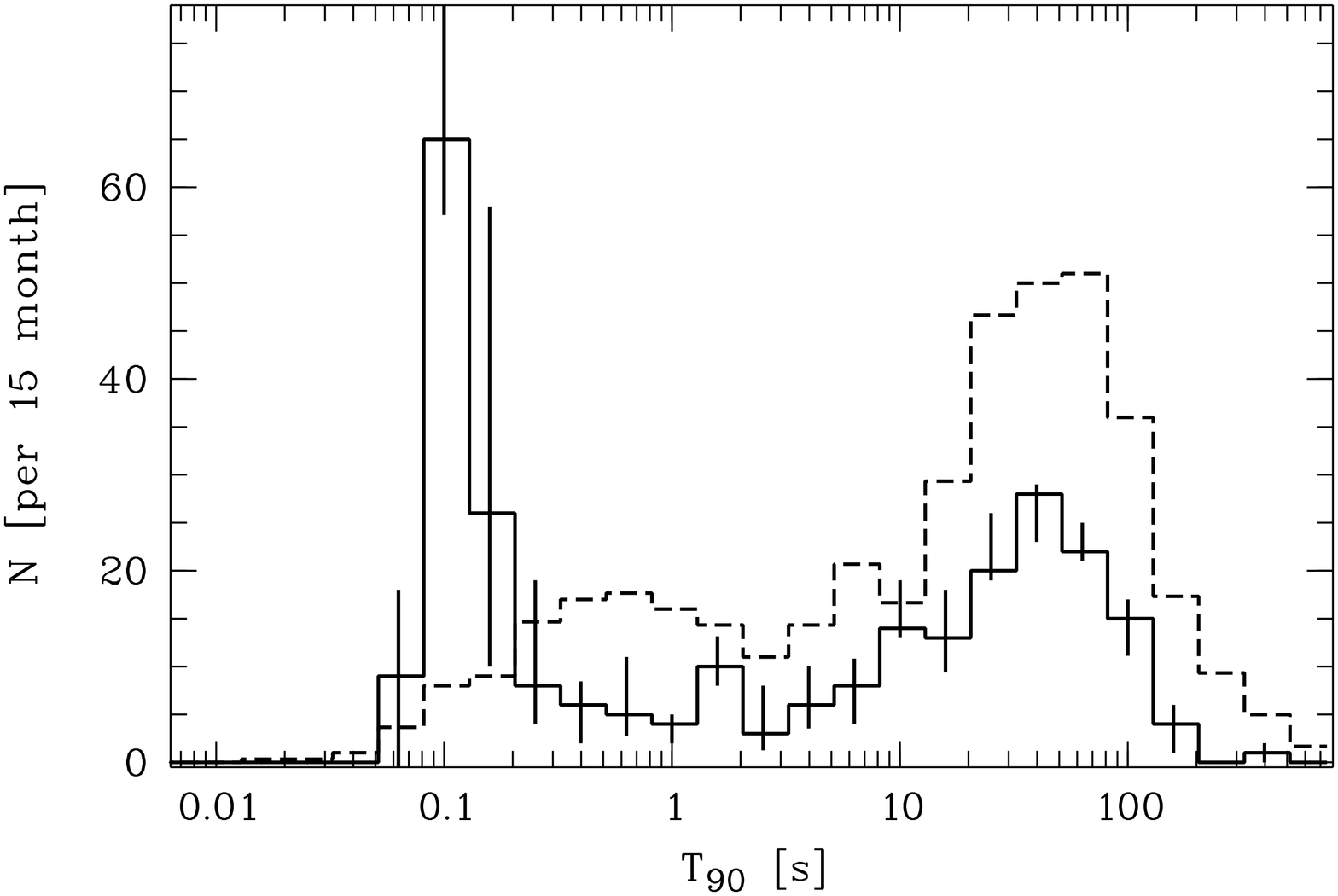} 
   \caption { Distribution of T$_{90}$ for all GRB candidates (solid
line) and for 1234 GRBs from the 4$^{th}$ BATSE GRB catalogue
(Paciesas \etal\ 1999; dashed). In order to compare with the SPI-ACS
detections, the BATSE distribution is scaled to the elapsed INTEGRAL
mission time (15 months). The known bimodality is present in our
sample, but note the very large fraction of short events compared to
the sample of BATSE bursts.}
         \label{fig:burstDurationAll} 
   \end{figure*}

For the observer frame duration of GRBs typically the T$_{90}$ value
is used. This measure corresponds to the time interval in which 90\%
of the burst emission (starting at 5\% and finishing at 95\%) is
observed. The distribution of T$_{90}$ for the sample of SPI-ACS burst
candidates is shown in Fig.~\ref{fig:burstDurationAll}. For comparison
the distribution derived from 1234 GRBs contained in the 4$^{th}$
BATSE GRB catalogue (Paciesas
\etal 1999), scaled to the elapsed mission time of INTEGRAL, is included.

Regardless of the low statistics, a bimodality in the distribution,
comparable to what was seen with BATSE, is found in the SPI-ACS
sample. A long-duration population with a maximum at $\sim$40\,s and a
short-duration population peaking at 0.1\,s are observed. While the
long-duration events resemble the BATSE results closely, a clear
deviation is detected for the short events. In the SPI-ACS sample the
short burst candidates are dominated by 0.1--0.15\,s events while for
BATSE the distribution was much more smooth and peaked at
0.6\,s. Also the ratio of short to long events is significantly
different. While the 4th BATSE catalogue included $\sim$70\% long and
$\sim$30\% short bursts, in the SPI-ACS sample the short burst
candidates make $\sim$50\% of the total events.  Also noticeable is the
peak of the distribution at 0.1\,s. Unfortunately, as this is close to the time
resolution of SPI-ACS, this population cannot be further
resolved.

\begin{figure*}
   \centering
   \includegraphics[width=0.6\textwidth,angle=270]{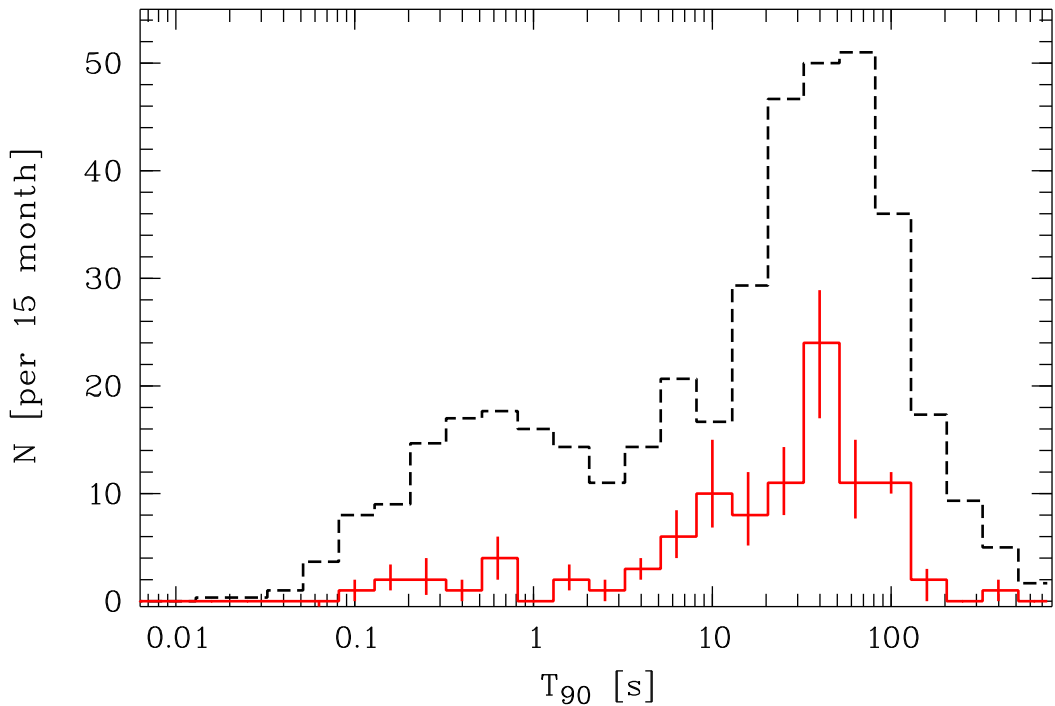} 
   \caption { Same as Fig.~\ref{fig:burstDurationAll} except that
     here only the sample of confirmed SPI-ACS bursts (solid line) is
     shown. A comparison with Fig.~\ref{fig:burstDurationAll} shows
   that only a small fraction of the short events was confirmed while
   nearly all long-duration bursts were also seen with other instruments.}
         \label{fig:burstDurationLoc} 
   \end{figure*}

In Fig.~\ref{fig:burstDurationLoc} we show the duration distribution
for the confirmed GRBs in the SPI-ACS sample. While $\sim$80\% of the
long-duration bursts are confirmed less than 6\% of the short duration
candidates were detected by other $\gamma$-ray
instruments. Although the small number statistics still do not allow a
quantitative statement to be made about the confirmed (short \& long) burst
distribution, it is similar to the BATSE results. Thus, the
question of the origin of the unconfirmed short population arises,
which will be discussed in the following section.

\section{origin of the short events}

There are various possible explanations for the significant excess of
unconfirmed 0.1--0.15\,s long events in the SPI-ACS sample. (i)
SPI-ACS observes a 'real' population of short very hard GRBs which has
gone undetected up to now due to the lack of sensitivity at very high
energies. The finding that the ratio of short to long burst candidates
is larger for SPI-ACS than for BATSE can for instance be explained
with the latter being sensitive in a lower energy range (50--320\,keV)
than SPI-ACS ($\sim$75\,keV to $>$10\,MeV).  However, several
$\gamma$-ray instruments which are now in orbit have the capability to
detect such hard events.  This was demonstrated for the event of
December 14 2003. Observations with the Gamma-Ray Spectrometer (GRS)
aboard Mars Odyssey demonstrated that this very bright short (0.3\,s)
burst had a very hard spectrum peaking at $\sim$2\,MeV (Hurley \etal
2003f). This event was simultaneously detected by Konus-Wind and
Helicon-Coronas-F (Golenetskii \etal 2003a, 2003b). So the fact the
majority of the SPI-ACS short events are only detected by the SPI-ACS,
but not by other instruments, argues against a new population. (ii) A
small contribution to the short burst population might also arise from
soft Gamma-ray repeaters (SGRs). SGRs are a small class (5 known
members) of objects that are characterized by brief (typical
$\sim$100\,ms), very intense ($\sim$10$^{44-45}$\,erg) bursts of hard
X-rays and soft gamma-rays. They are generally located near the
Galactic plane.  It is likely that they originate from strongly
magnetized neutron stars, or magnetars (Duncan \& Thompson 1992).  As
the light curve shapes are similar, events in the SPI-ACS rate
originating from SGRs cannot be easily distinguished from short GRBs
without the localization by triangulation with other
instruments. (iii) A significant contribution to these short events
comes from instrumental effects and/or cosmic-ray events. This cannot
be ruled out.

\begin{figure}[h] \centering
   \includegraphics[width=0.5\textwidth]{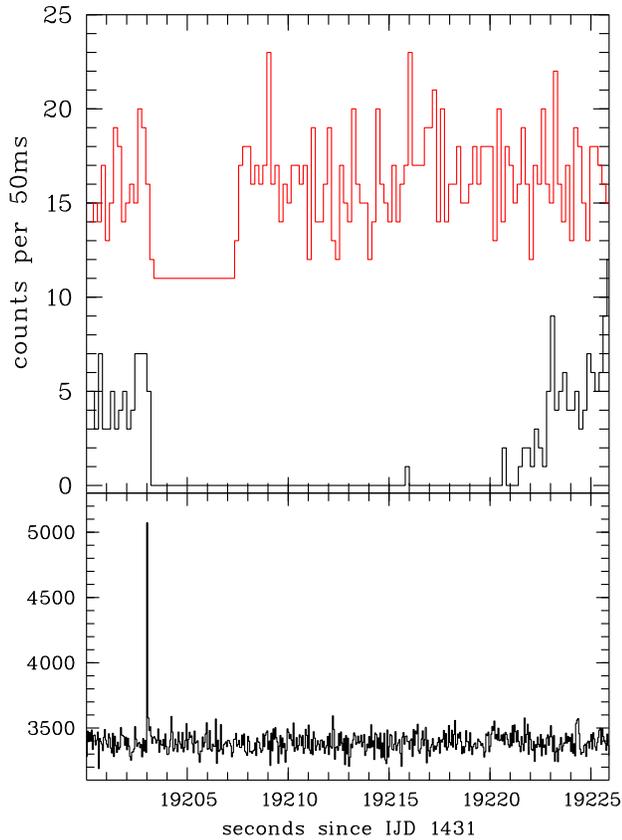} 
   \caption {Short
   event on December 1 2003 with simultaneous saturation in the
   Ge-detectors \#9 (upper) and \#10 (lower). The upper curve
   (detector \#9) is offset for better visibility.}  
   \label{fig:1431}
   \end{figure}

A detailed study of the population of short events revealed that a
significant fraction is accompanied by the simultaneous saturation of
one or several Ge-detectors of the spectrometer. An example for an
event on December 1 2003 is shown in Fig.~\ref{fig:1431}. At the time
of the significant rate increase in SPI-ACS over $\sim$50\,ms the
Ge-detectors \#9 and \#10 went into saturation and stayed there for
$\sim$4 and $\sim$20\,s, respectively. A statistical analysis of the
data of the mission shows that a saturation event occurs approximately
every four hours and that nearly all of these events have a
simultaneous rate increase in SPI-ACS. An independent search for short
events in the SPI-ACS overall count rate found $\sim$15 events per
hour at a 4.5$\sigma$ level. Thus approximately every 60$^{th}$ short
SPI-ACS event coincides with a saturation in the Ge-detectors. Note,
the chance coincidence of the SPI-ACS events happen at the same time
as the saturation in the Ge-detectors is $\sim$3$\times$10$^{-8}$,
which makes the causal connection evident.

These observations suggest a particle origin for at least a fraction
of the short burst candidates. The most natural explanation for the
saturation of the Ge-detectors is the deposit of a large energy amount
in the crystals. A very energetic cosmic-ray particle can hit a BGO
crystal and deposit part of its energy, thus producing the short count
rate increase (see below for a discussion). Depending on the flight
direction, the particle can further hit one or several Ge-detectors
and cause the saturation, or pass through the BGO for a second time,
or not interact again. The probability of a particle hitting the
SPI-ACS and Ge-detectors can be estimated from the geometry of the
instrument to be $\sim$1/40 of the probability to pass only through
the anticoincidence shield. This is in rough agreement with the rate
of SPI-ACS short events that coincide with saturations. Therefore,
from this simple assumption one can conclude that at least a
significant fraction of the unconfirmed short burst candidates can
originate from cosmic-ray particles hitting the instrument. Comparing
the rate of events in the SPI-ACS with the observed rate of
cosmic-rays as a funtion of energy (Cronin \etal 1997), the energy for
a single particle can be estimated to be $\sim$300\,GeV and above.

Most of the saturation events occur in a single Ge-detector. Hits in
several detectors are significantly less frequent. If the saturation
is caused by a single particle then a particle passing through the
coded mask will only deposit energy in one detector as the
cross-section for hitting more than one Ge-crystal is very low. A
particle hitting the detector plane with an angle not perpendicular
to the detector plane can deposit energy (and cause
saturation) in several detectors. That is in agreement with the
observation that saturation events in more than one detector are
always accompanied by a clear SPI-ACS count rate increase, while a
small fraction of the saturation events in single detectors show no
corresponding signal in the SPI-ACS overall rate. 

In the case of a particle shower the number of detector hits depends
on the width of the distribution of the secondary particles and on the
inclination angle.  In IBIS/PICsIT strong bursts (2500 counts) of up
to 170\,ms of particle-induced showers are observed (Segreto
2002). These showers are thought to be produced by cosmic-ray hits on
the satellite or detector structure. As a consequence of the
interaction of the primary particle with matter, high-energy photons
and electron-positron pairs are created. These in turn produce a
cascade of secondary electrons and photons via bremsstrahlung and pair
production. The time profiles of the events observed in PICsIT are
rather complex and both straight tracks and closed areas are visible
(Segreto 2002). These observations are similar to what is observed
with the saturation in SPI and thus support the idea of cosmic-rays as
an origin of the short SPI-ACS events.

The assumption was made above that the passage of a very energetic
particle produces the observed short increase of the SPI-ACS overall
rate. One possibility is that the deceleration of the particle in the
BGO crystal produces a long-lasting (50-150\,ms) phosphorescent
afterglow energetic enough to produce recurrent triggers of the
electronics. BGO was originally selected for SPI-ACS because of its
short decay times and very low phosphorescence compared to
e.g. CsI(Na) and NaI(Tl). It has decay times of 60\,ns and 300\,ns
where the second is by far the more probable (90\% vs. 10\% for the
60\,ns) and additionally a so-called afterglow (not to be confused
with the GRB afterglow) of 0.005\% is expected at 3\,ms. The afterglow
originates from the presence of millisecond to even hour long decay
time components and is believed to be intrinsic and correlated to
certain lattice defects.

In order to produce a single count in the SPI-ACS rate, a minimum of
75\,keV (the lower energy threshold) has to be accumulated during an
integration time of 600\,ns. With a total light yield for BGO of
8000-10000 photons/MeV, this corresponds to $\sim$750
photons. Assuming a typical decay law of I(t)=I$_0$*e$^{-t/\tau}$
shows that the radiation will decrease to 0.005\% at 3\,$\mu$s. Thus
the decay is so fast that only a small number of counts ($\sim$3) will
be produced. From the afterglow properties the minimum original
excitation energy can be estimated. For instance, to have 1000
recurrent triggers (a count rate increase of 1000) in a single
crystal, the afterglow must be bright enough to be above the 75\,keV
threshold for $\sim$3\,ms (integration time + dead time). The
0.005\,\% afterglow emission at 3\,ms thus corresponds to a
$\sim$1.6\,GeV initial excitation using the photon yield given
above. Therefore, particles which deposit $\sim$1.6\,GeV or more in a
crystal can indeed be the origin of the short event population in
SPI-ACS. Short events showing similar temporal behaviour were already
discussed for CsI(Na) and CsI(Tl) crystals exposed to primary cosmic
radiation (Hurley 1978). The same conclusion for the origin, namely
cosmic-ray nuclei in the iron group, was drawn. Note that CsI has a
significantly more intense afterglow (0.1--1\,\% at 6\,ms).




\section{Conclusion}

It is now clear that the prominent population of short burst
candidates detected by SPI-ACS is probably strongly contaminated by
cosmic-ray events. At present it is not feasible to disentangle the
true short GRBs, with durations $<$0.2\,s, from these events without a
confirmation by other $\gamma$-ray instruments. Therefore, we can only
give the approximate rate of GRBs observed by SPI-ACS with durations
$>$0.2\,s to be $\sim$140 (88 confirmed by other instruments) per
year.



\begin{thebibliography}{}
\bibitem[Costa:1997]{Costa:1997} Costa E., Frontera F., Heise J. \etal 1997, Nature 387, 783
\bibitem[Cronin:1997]{Cronin:1997} Cronin J., Gaisser T.K. \& Swordy
S.P. 1997, Sci. Amer. 276, p44 
\bibitem[Duncan:1992]{Duncan:1992} Duncan R.C. \& Thompson C. 1992, ApJ 392, L9
\bibitem[Fishman:1995]{Fishman:1995} Fishman G.J., Meegan C.A., Wilson
R.B., \etal 1989, in Proc. of the GRO Science Workshop, GSFC,2 
\bibitem[Golenetskii:2003a]{Golenetskii:2003a} Golenetskii S., Mazets E., Pal'shin V. \& Frederiks D. 2003a, GCN 2487
\bibitem[Golenetskii:2003b]{Golenetskii:2003b} Golenetskii S., Mazets E., Pal'shin V. \& Frederiks D. 2003b, GCN 2488
\bibitem[Hjorth:2003]{Hjorth:2003} Hjorth J., Sollerman J., M{\o}ller P. \etal 2003, Nature 423, 847
\bibitem[Hurley:1978]{Hurley:1978} Hurley K. 1978, A\&A 69, 313
\bibitem[Hurley:1997]{Hurley:1997} Hurley K. 1997, in Proc. of the 2nd
INTEGRAL Workshop, ESA SP 382, 491
\bibitem[Hurley:2003a]{Hurley:2003a} Hurley K., Cline T., von Kienlin A., Lichti G.G. \& Rau A. 2003a, GCN 1874
\bibitem[Hurley:2003b]{Hurley:2003b} Hurley K., Mazets M., Golenetskii S., \etal\ 2003b, GCN 1937
\bibitem[Hurley:2003c]{Hurley:2003c} Hurley K., Cline T., Smith D.M. \etal\ 2003c, GCN 2281
\bibitem[Hurley:2003d]{Hurley:2003d} Hurley K., Cline T., Mitrofanov I. \etal\ 2003d, GCN 2378
\bibitem[Hurley:2003e]{Hurley:2003e} Hurley K., Cline T., Mitrofanov I. \etal\ 2003e, GCN 2441
\bibitem[Hurley:2003f]{Hurley:2003f} Hurley K., Cline T., Mazets E. \etal\ 2003f, GCN 2448
\bibitem[Hurley:2004]{Hurley:2004} Hurley K., Rau A., von Kienlin A. \& Lichti G.G. 2004, these proceedings
\bibitem[Kienlin:2003]{Kienlin:2003} von Kienlin A., Beckman V., Rau
A. \etal\ 2003, A\&A 411, L299
\bibitem[Klebesadel:1973]{Klebesadel:1973} Klebesadel R., Strong I. \&
Olsen R. 1973, ApJ 182, L85
\bibitem[Lichti:2000]{Lichti:2000} Lichti G.G., Georgii, R., von Kienlin A. \etal\ 2000, in Proc. of the 5th Compton Symp., AIP Conf. Proc. 510, 722
\bibitem[Mereghetti:2004]{Mereghetti:2004} Mereghetti S. 2004, these proceedings
\bibitem[Paciesas:1999]{Paciesas:1999}  Paciesas W.S., Meegan C.A., Pendleton G.N. \etal\, 1999, ApJS 122, 465
\bibitem[Ryde:2003]{Ryde:2003} Ryde F., Borgonovo L., Larsson S., Lund N., von
  Kienlin A. \& Lichti G.G. 2003, A\&A 411, L331
\bibitem[Segreto:2002]{Segreto:2002} Segreto A., 2002, IN-IB-IASF/Pa-RP-030/02
\bibitem[Stanek:2003]{Stanek:2003} Stanek K. Z., Matheson T., Garnavich P. M. \etal 2003, ApJ 591, L17
\end{thebibliography}
\end{document}